# Variation of the vortex activation energy, $U(T, H)$, with hole content in $YBa_2Cu_3O_{7-\delta}$ thin films

**S H Naqib and R S Islam**

*Department of Physics, University of Rajshahi, Rajshahi-6205, Bangladesh*

E-mail: salehnaqib@yahoo.com

**Abstract**

The nature of in-plane resistive transition of high-quality *c*-axis oriented crystalline thin films of $YBa_2Cu_3O_{7-\delta}$ have been studied under magnetic fields ($H$) applied along the *c*-direction over a wide range of doped holes, $p$, in the $CuO_2$ planes. The field and temperature dependent in-plane resistivity, $\rho_{ab}(T, H)$, below the mean field superconducting transition temperature, $T_p$, has been analyzed within the thermally assisted flux-flow (TAFF) scenario. We have extracted the temperature and field dependent flux activation energy, $U(T, H)$ from this analysis. The low-$T$ part of the $\rho_{ab}(T, H)$ data can be described quite well by a dimensionless activation energy having the functional form of $U(T, H) = (1-t)^m (H_0/H)^{-\beta}$, where $t = T/T_p$, is the reduced temperature and $H_0$, is a characteristic field scale that primarily determines the magnitude of the activation energy for a given sample composition. The temperature exponent, $m$, shows a systematic variation with the hole content, whereas the field exponent, $\beta$, is insensitive to the $p$-values and is close to unity irrespective of the film composition. $H_0$, on the other hand, changes rapidly as hole concentration is varied. Possible implications of these results are discussed in this paper.

## 1. Introduction

Since the early days of their discovery, the peculiar broadening of the resistive transition under applied magnetic field in high-$T_c$ cuprates has attracted a lot of attention [1 – 3]. The phenomenon is particularly interesting because of its link to magnetic vortex dynamics. The mechanism of pinning and the motion of the flux lines are the most important fields of applied superconductivity research since they determine the magnitude as well as the temperature and field dependences of the critical current density. The various theoretical interpretations for the experimentally observed behavior of the temperature and field dependent resistivity, $\rho(T, H)$, at the vicinity of superconducting transition are still not entirely satisfactory. Complications arise because of the electronic and structural anisotropy, low superfluid density, small coherence volume, and high thermal energy associated with high-$T_c$ itself, in all the families of hole doped cuprates. These features enhance the role of pairing fluctuations and at the same time weaken the flux pinning properties. Generally, over a certain range of temperature, sufficiently below the mean field $T_c$ ($\equiv T_p$, in this paper), where critical fluctuations are believed to be absent, the resistive features are modeled using the thermally assisted flux flow (TAFF) formalism [4 – 8]. This is the regime where the *vortex-liquid* exhibits viscous flow.

As far as the potential for future large scale applications are concerned, YBa$_2$Cu$_3$O$_{7-\delta}$ (Y123) remains as the most promising system because of its high irreversibility field, low structural anisotropy, and relatively higher superfluid density (which results in large intrinsic depairing critical current density). The normal and superconducting (SC) state properties of all the cuprates depend strongly on the number of doped holes in the CuO$_2$ plane. Previously we have shown that the in-plane critical current density is primarily governed by the depairing contribution and therefore shows the same qualitative trend as shown by the *p*-dependent superfluid density and condensation energy [9, 10]. We have also detected a crossover in the temperature dependence of the critical current density in the underdoped (UD) Y123 where depinning contribution takes over [11]. To obtain a comprehensive picture, we have undertaken this project to investigate the temperature and field dependences of the activation energy, $U(T, H)$, as a function of hole content in high-quality *c*-axis oriented thin films of Y123. It is rather surprising that even though a number of work on $U(T, H)$ already exist for Y123 and related systems [4 – 8, 12 – 14], most of these studies were done at a single sample composition, or by considering the role of sample anisotropy alone. So far, no systematic study has been done on the nature of the *p*-dependence of $U(T, H)$ in Y123 explicitly, to the best of our knowledge. From the detailed analysis of the in-plane resistivity, $\rho_{ab}(T, H)$, data we have found that the dimensionless activation energy can be expressed as $U(T, H) = (1-t)^m (H_0/H)^{-\beta}$, irrespective of the hole content. Here $t = T/T_p$ is the reduced temperature and $H_0$ is a characteristic field scale. The value of the exponent *m* depends on the hole content, whereas $\beta$ is insensitive to the *p*-values and is close to unity. $H_0$, on the other hand, decreases rapidly as the hole concentration is reduced. We have discussed the implications of these results in the concluding section of this paper.

## 2. Sample preparation and characterization

Crystalline *c*-axis oriented thin films of $YBa_2Cu_3O_{7-\delta}$ were synthesized from high-density single-phase sintered targets using the method of pulsed laser deposition (PLD). Thin films were grown on (*001*) $SrTiO_3$ single crystal substrates at a deposition temperature of 780°C under an oxygen partial pressure of 0.95 mbar. Samples were characterized by using X-ray diffraction (XRD), atomic force microscopy (AFM), *ab*-plane room-temperature thermopower, $S_{ab}$[*290 K*], and $\rho_{ab}(T)$ measurements. XRD was used to determine the structural parameters, phase-purity, and degree of *c*-axis orientation (from the rocking-curve analysis). AFM was employed to study the grain size and the thickness of the films. All the films used in the present study were phase-pure and had high-degree of *c*-axis orientation (typical value of the full width at half-maximum of (*007*) peak was ~ 0.25°). Thickness of the films lies within the range (2800 ± 300) Å. The hole concentrations were changed by changing the oxygen deficiency in the $CuO_{1-\delta}$ chains by annealing the films under different temperatures in different oxygen environments. $S_{ab}$[*290 K*] was used to calculate the planar hole concentration following the method proposed by Obertelli *et al.* [15]. The level of oxygen deficiency was determined from an earlier work where the relation between $\delta$ and *p* were established [10]. Also, as an independent check, systematic changes in the *c*-axis lattice parameters were noted as $\delta$ changes as a result of oxygen annealings [10]. $\rho_{ab}(T)$ measurements gave information regarding the impurity content and about the quality of the grain-boundaries of the films. All the films used in this study have low values of $\rho_{ab}(300\ K)$ and the extrapolated zero temperature resistivity, $\rho_{ab}(0\ K)$. Details of the PLD method and film characterizations can be found in refs. [16, 17]. Fig. 1 shows the XRD profile of the as-prepared thin film annealed *in-situ* at 400° C in 100% oxygen environment. We have used thin films with four different hole contents in this study. The *p*-values for these films are 0.170 (OD), 0.150 (OPD), 0.135 (UD1), and 0.118 (UD2), where the quoted values are accurate within ± 0.005.

## 3. Experimental results

The resistivity, under applied magnetic fields along the *c*-direction, was measured using the four-terminal configuration on patterned untwined thin film samples. Evaporated gold pads were used to form low resistance contacts. An AC current of 10μA (77Hz) was applied. Low noise transformers were used to improve the signal-to-noise ratio, and voltages were measured with *EG & G model 5110* lock-in amplifiers. Measurements were taken at magnetic fields of 0, 3, 6, 9 and 12 Tesla in the temperature range from 50 – 160 K (the upper temperature limit for the zero-field and the 12 Tesla data was 300 K). A temperature sweep rate of 2 K/min was used. A 15-Tesla magnet system (*Oxford Instruments*) was used to carry out these magneto-transport measurements. A *Lake Shore 340* temperature controller was used to control the temperature of the probe head. Temperature was monitored via a *Cernox* thermometer placed at close proximity of the sample. $\rho_{ab}(T, H)$ data for all the thin films are shown in Figs. 2.

## 4. Analysis of the $\rho_{ab}(T, H)$ data

Within the context of the TAFF model the single most important parameter is the activation energy for the motion of flux lines/bundles. There exists a number of methods to extract the temperature and field dependent $U(T, H)$ from the analysis of the resistivity data [4 – 8, 12, 18]. In this paper we have adapted the method used earlier by Liu *et al.*, [8]. This method has been checked for internal consistency and provides us with a satisfactory scheme for the analysis of the $\rho_{ab}(T, H)$ data below $T_p$, where the critical fluctuations are absent and a simple Arrhenius equation describes the resistive broadening under applied magnetic field. In the TAFF region the field dependent resistivity is expressed as follows

$$\rho(T,H) = \rho_n \exp[-U(T,H)/k_B T] \qquad (1)$$

where $\rho_n$ is the normal state resistivity. The activation energy depends both on temperature and the magnetic field and can be expressed as the product of two functions in the dimensionless form as $U(T, H) = f(T)h(H)$ [7, 8]. In this case, $U(T, H)$ can be thought of as the activation energy normalized by $U_0$, where $U_0$ is the effective unperturbed pinning potential at $T = 0$ K. The functions $f$ and $h$ have the following forms [7, 8]: $f(T) = (1-t)^m$ and $h(H) = (H_0/H)^{-\beta}$. Based on a scaling analysis, Yeshurun and Malozemoff [19] suggested $m = 1.5$, for cuprate superconductors. Subsequent experimental analyses found that the value of $m$ varies over a wide range for different families of cuprates. Even for the same system, depending on the method of synthesis and crystalline state, different values of $m$ were obtained [7, 8]. As mentioned previously, ref. [8] provides us with a way to calculate the value of $m$ without any prior assumption. The central feature of this procedure is the observation that over a significant temperature range below $T_p$, it is possible to describe $f(T)$ by a function $g(T)$ of the form given by

$$g(T) = (T_p - T)^m \qquad (2)$$

At the same time, following refs. [8, 12, 13], we make use of the fact that the applied magnetic field and the zero-resistance temperature, $T_{c0}$, can be related through the equation expressed as

$$H = H_0 \left[\frac{1 - T_{c0}(H)/T_c}{T_{c0}(H)/T_c}\right]^{-\beta} \qquad (3)$$

Here $T_c$ denotes the SC transition temperature in the absence of any magnetic field, i.e. $T_c \equiv T_p$ for the field-free case. In cuprate superconductors the mixed state, significantly below the mean field SC transition temperature, can be described as a truly vortex solid state, whereas the state at temperatures closer to $T_p$ is dissipative and the vortices behave as liquids even for a very low current density. In ultra-clean compounds the two regimes are separated by a well-defined vortex lattice melting temperature [20, 21]. In samples with quenched disorders, the solid-liquid transition becomes glass-like and $T_{c0}$ can be considered as the glass transition temperature, i.e., at $T < T_{c0}$, the vortices are pinned and are in the non-dissipative glassy state. Eq. 3 describes the field induced shifts in $T_{c0}$ quite well. Fits to the $T_{c0}(H)$ data gave the value of $\beta$ close to unity ($0.98 \pm 0.04$), in agreement with previous studies [7, 8], irrespective of the hole content. These fits also gave estimates for $H_0$ as hole contents vary. Using Eq. 3, the activation energy can now be expressed in terms of $T_{c0}$ as follows

$$U(H,T) = \frac{f(T)}{[(T_c - T_{c0}(H)/T_{c0}(H)]} \tag{4}$$

Therefore, Eq. (1) turns into

$$\rho(T,H) = \rho_n \exp[-\frac{f(T)T_{c0}(H)/((T_c - T_{c0}(H))}{k_B T}] \tag{5}$$

Rearranging the above expression, we obtain the following expression for the function $f(T)$ in terms of experimentally measurable parameters

$$f(T) = -\frac{k_B T[T_c - T_{c0}(H)]}{T_{c0}(H)} \ln[\frac{\rho(T)}{\rho_n}] \tag{6}$$

At this point, it should be noted that the above equation contains a number of parameters (namely, $T_c$ ($\equiv T_p$), $T_{c0}$, and $\rho_n$) which are determined somewhat subjectively [4 – 8, 12 – 14]. Different methods yield slightly different values [4 – 8, 12 – 14] but the main conclusions do not differ significantly irrespective of the criterion used to locate these parameters. In this study, we have defined $T_p$ as the temperature where $d\rho_{ab}(T, H)/dT$ is maximum and $T_{c0}$ has been taken at the temperature where $\rho_{ab}(T, H)$ goes to zero. The normal state resistivity was located at the intersection between two straight lines, one drawn as a tangent to the $\rho_{ab}(T, H)$ curve at $T_p$ and the other one as the extrapolated straight line fitted to the $\rho_{ab}(T, H)$ data over a $T$-range from $T_p + 50$ K to $T_p + 25$ K. The reasons for selecting this particular temperature range are the followings – (i) Ideally, $\rho_n$ should be the resistivity just above at the SC transition temperature in the absence of any SC fluctuation. Previous studies on different families of cuprates showed that significant pairing fluctuations only start to contribute in paraconductivity from around 20 K above the mean field SC transition temperature [22 – 24]. Therefore, $T$-region too close to $T_p$ has been avoided and (ii) Higher temperature linear extrapolation has been excluded since the pseudogap (PG) [22, 25] induces a downturn from the high-$T$ linearity in the resistivity data in the normal state and this contribution due to the PG should be included in the extracted value of $\rho_n$ as PG probably originates from a normal state correlation [22]. The procedures for determining $T_p$, $T_{c0}$, and $\rho_n$ are illustrated in Figs. 3a and 3b. The calculated values of $f(T)$ using Eq. 6, is shown in Fig. 4. We have also shown the fitted function $g(T)$, given by Eq. 2 in the same plot. It is seen that $g(T)$ describes the $f(T)$ data quite well at temperatures below $T_p$, in agreement with ref. [8]. Significant order parameter fluctuations are expected close to $T_p$ and the Eq. 1 holds only in the TAFF region, therefore it is somewhat expected that a single exponent fit does not describe the experimentally obtained $f(T)$ over the entire $T$-range from $T_p$ to $T_{c0}$. According to the procedure employed here, the plot of $\ln(\rho/\rho_n)$ vs. $[(T_p - T)^m (H_0/H)^\beta]/T$ should be a straight line and all the data points at different $p$ values could be scaled onto a single straight line. These predictions were checked and our data satisfied such scaling analysis over an extended $T$-range, except at temperatures close to $T_p$ below the SC transition. The values of the temperature exponent, $m$, obtained from this analysis are not sensitive (within the error bars) to the magnitude of the applied magnetic field in agreement with previous studies [7, 8] but rather show a systematic variation with the hole content. We have shown these behaviors in Fig. 5. Fig. 6 shows the strong dependence of $H_0$ on the hole concentration.

Once reliable estimates for $m$, $\beta$, and $H_0$ are made, the field and temperature dependent activation energy can be determined uniquely. We have extracted the field dependent part of the activation energy, $U(H)$, from the slope of the $\ln(\rho/\rho_n)$ vs. $(T_p - T)^m/T$ plot. The result of this analysis is illustrated in Fig. 7. It is observed that $U(H)$ varies almost linearly with $1/H$. The slope of these lines at different $p$ gives an independent measure of $H_0(p)$. The values of $H_0(p)$ obtained from these slopes agree very well (within 5%) with those shown in Fig. 6, extracted from the fitting of the $T_{c0}(H)$ data using Eq. 3.

## 5. Discussion and conclusions

As stated earlier, previous studies have linked the values of $m$ to extrinsic effects such as the method of synthesis, nature of pinning centers and to intrinsic properties like the particular family of cuprates or to the degree of electronic anisotropy [7, 8, 26]. Here, we have found clear indications that for a given system $m$ varies systematically with hole content in the $CuO_2$ planes. Quite interestingly, it has been suggested earlier that $m = 1.5$ and 2.0 corresponds to 3D and 2D behaviors, respectively [26]. Our study reveals that $m$ decreases with the decrease in the hole content. This trend is completely opposite to what was suggested in ref. [26], since the films become more and more anisotropic as $p$ decreases. For all the films under study, $β$ remains close to unity. This could imply [26, 27] that the flux pinning centers are predominantly point defects in these untwined thin films.

The values of $H_0$ are strongly $p$-dependent. $H_0$ decreases rapidly in the UD region and seems to peak in the overdoped (OD) side. It is interesting to note that even though the $T_c$ values are almost identical for the OPD and the OD samples, $H_0$ is significantly higher for the OD compound. The OPD compound is more disordered due to oxygen deficient $CuO_{1-δ}$ chains, whereas the OD one has a higher superfluid density due to a smaller PG in the low-energy electronic density of states [22, 28]. These imply that the role played by superfluid density/SC condensation energy is significantly greater in enhancing the pinning potential than that due to the oxygen defects in the $CuO_{1-δ}$ chains. This is consistent with the collective pinning model [29], where the condensation energy is directly proportional to the pinning potential. It is perhaps noteworthy that both condensation energy and superfluid density increase the depairing critical current density and these two intrinsic SC parameters depend largely on the magnitude of the PG. Therefore, it appears that to maximize the critical current density or the irreversibility field, one must control the sample composition so that the PG is small or zero, without reducing $T_c$ too much by gross overdoping. It is also interesting to note that $p$-values of the two UD films are quite close to the 1/8[th] doping where the spin/charge stripe correlations are at their strongest [30, 31]. This study shows that the exponents $m$ and $β$, as well as the characteristic field $H_0$, vary smoothly as a function of hole content and no noticeable extra feature appears in the vicinity of the 1/8[th] *anomaly*. This indicates that the inhomogeneity in the charge/spin arrangements induced by the incipient stripe instability does not affect the vortex dynamics significantly in Y123.

To summarize, we have explored the nature of *ab*-plane resistive transition of high-quality *c*-axis oriented crystalline thin films of YBa$_2$Cu$_3$O$_{7-\delta}$ over a wide range of hole concentrations. We have found that the temperature exponent varies systematically with hole content. This exponent becomes smaller as the hole content decreases (as the system becomes more anisotropic). The field exponent, on the other hand, remains largely unaffected. The characteristic magnetic field $H_0$, that primarily determines the pinning strength, decreases sharply with decreasing $p$ in the UD side. It appears that $H_0$ is controlled by the SC condensation energy/superfluid density and therefore, is linked to the PG in the quasiparticle energy spectrum.

**Acknowledgements**

The authors would like thank the MacDiarmid Institute for Advanced Materials and Nanotechnology, New Zealand, and the IRC in Superconductivity, University of Cambridge, UK, for funding this research. The authors would also like to thank the AS-ICTP, Trieste, Italy, for the hospitality.

**Figure captions**

Figure 1: X-ray diffraction (XRD) profile (obtained with Co $K_\alpha$ radiation) of the OD film. The Miller indices (00*l*) are shown in the plot.

Figure 2 (color online): The *ab*-plane resistivity of (a) OD, (b) OPD, (c) UD1, and (d) UD2 thin films under different magnetic fields applied along the crystallographic *c*-direction.

Figure 3 (color online): Representative plots for the OD sample in zero magnetic field, illustrating the methods employed to obtain (a) $T_{c0}$ and $T_p$, and (b) $\rho_n$ (see text for details) from the resistivity data.

Figure 4 (color online): $f(T)$: circles for the OD and squares for the UD2 thin films. $g(T)$ (dashed lines) is the low-$T$ fits to $f(T)$ (see text for details) at different applied magnetic fields. For clarity, data for two out of the four different film compositions are shown here.

Figure 5 (color online): Main: Variation of the exponent *m* with applied magnetic field for samples with different hole contents. *p*-values shown in the plot are accurate within ± 0.005. Inset: The average values of *m* versus *p*. The dashed curve is drawn as a guide to the eyes.

Figure 6 (color online): The characteristic field $H_0$ versus *p*. The dashed line is a second-order polynomial fit to the $H_0(p)$ data.

Figure 7 (color online): The dimensionless activation energy, $U(H)$ versus $1/H$ at different hole concentrations. The *p*-values given in the plot are accurate within ± 0.005. The dashed lines are fits to straight lines passing through the origin.

Figure 1

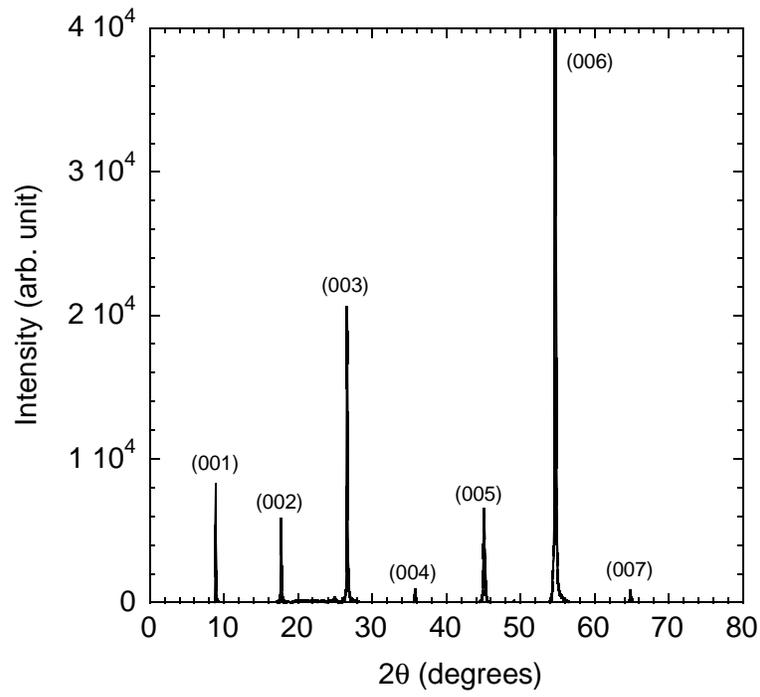

Figure 2

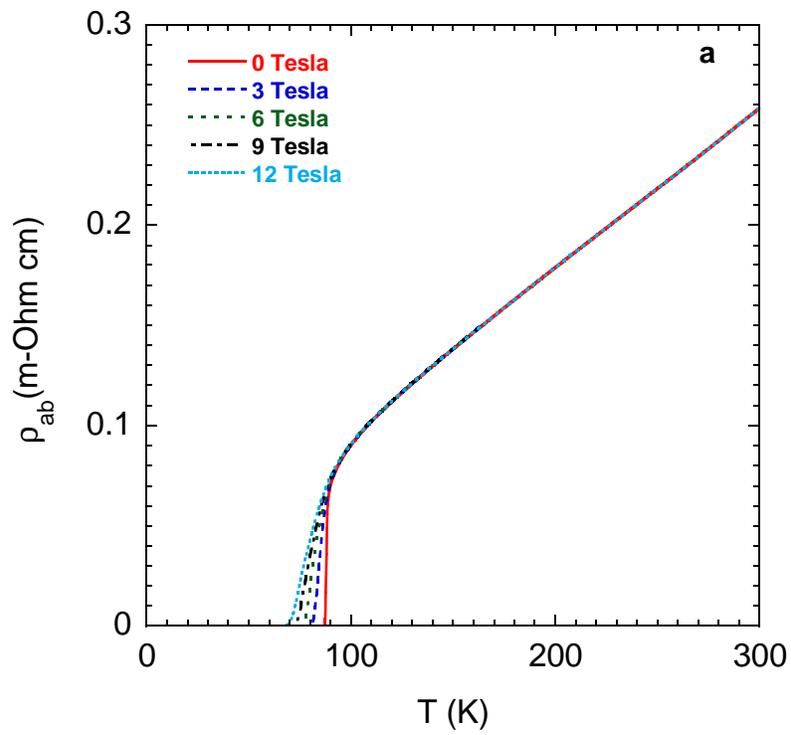

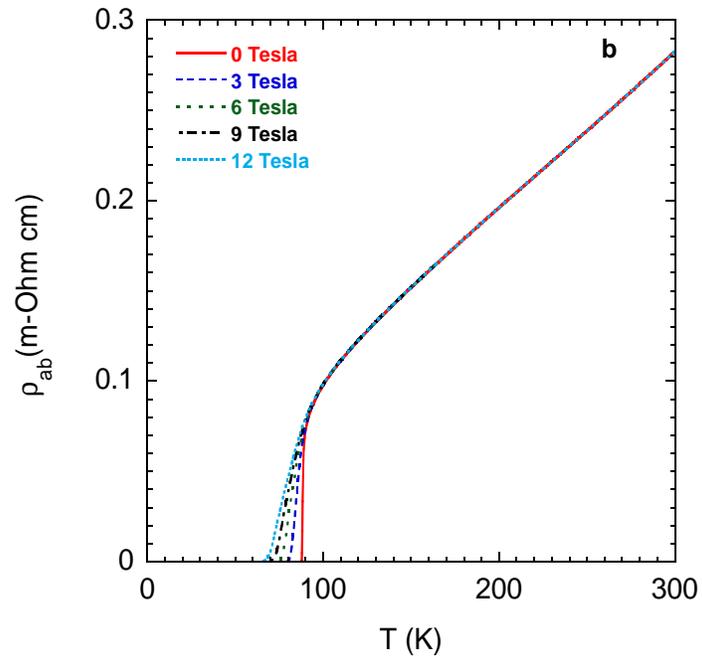
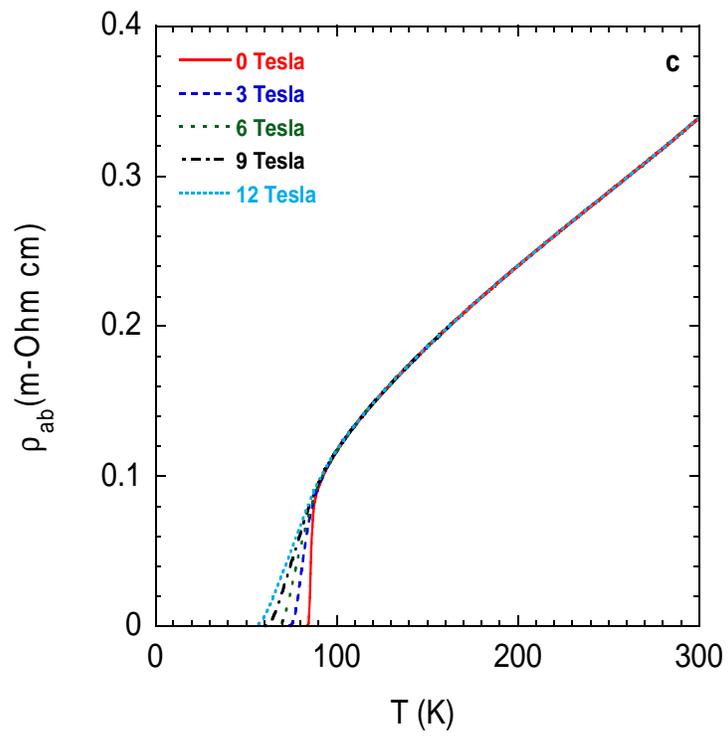

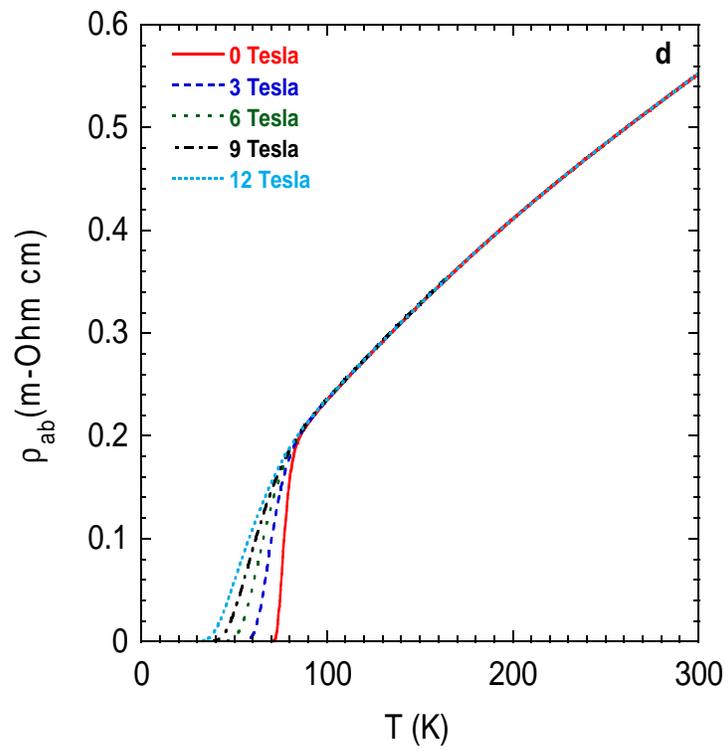

Figure 3

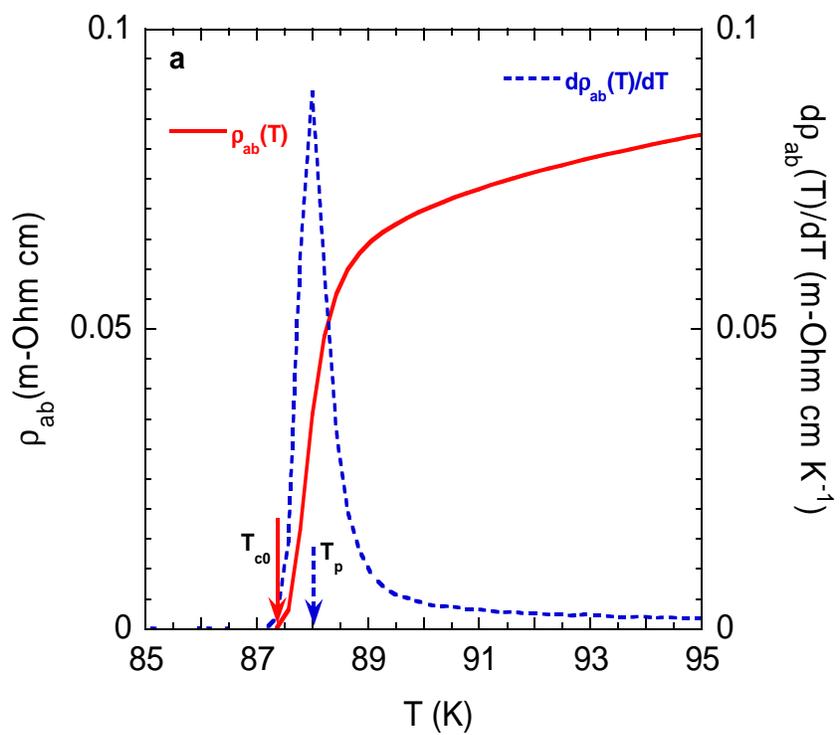

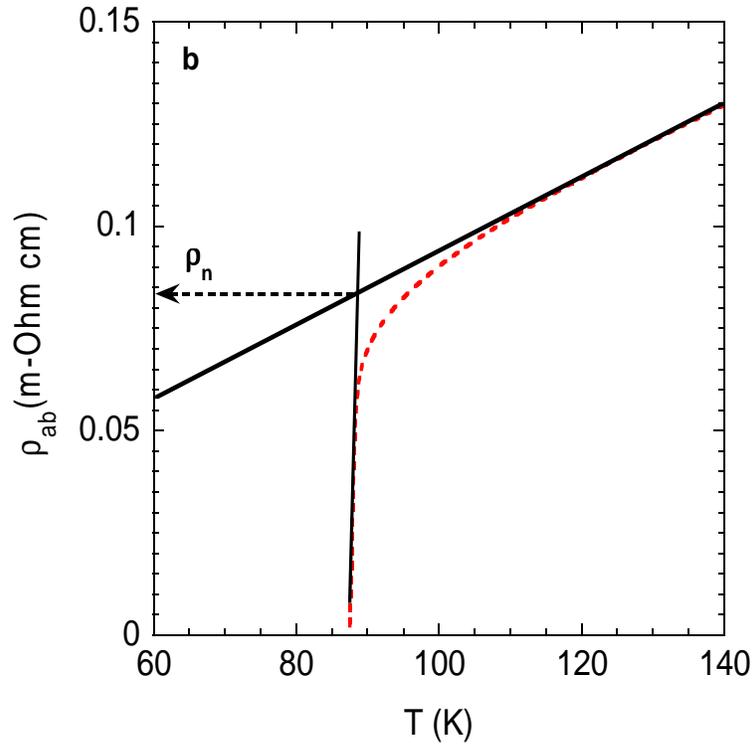

Figure 4

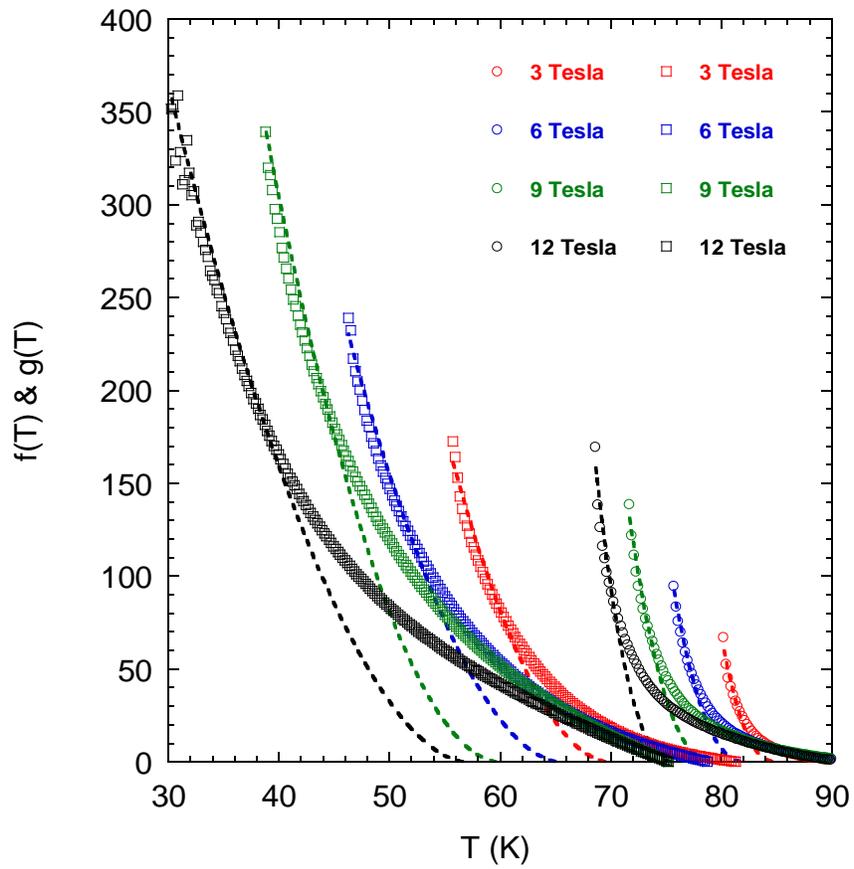

Figure 5

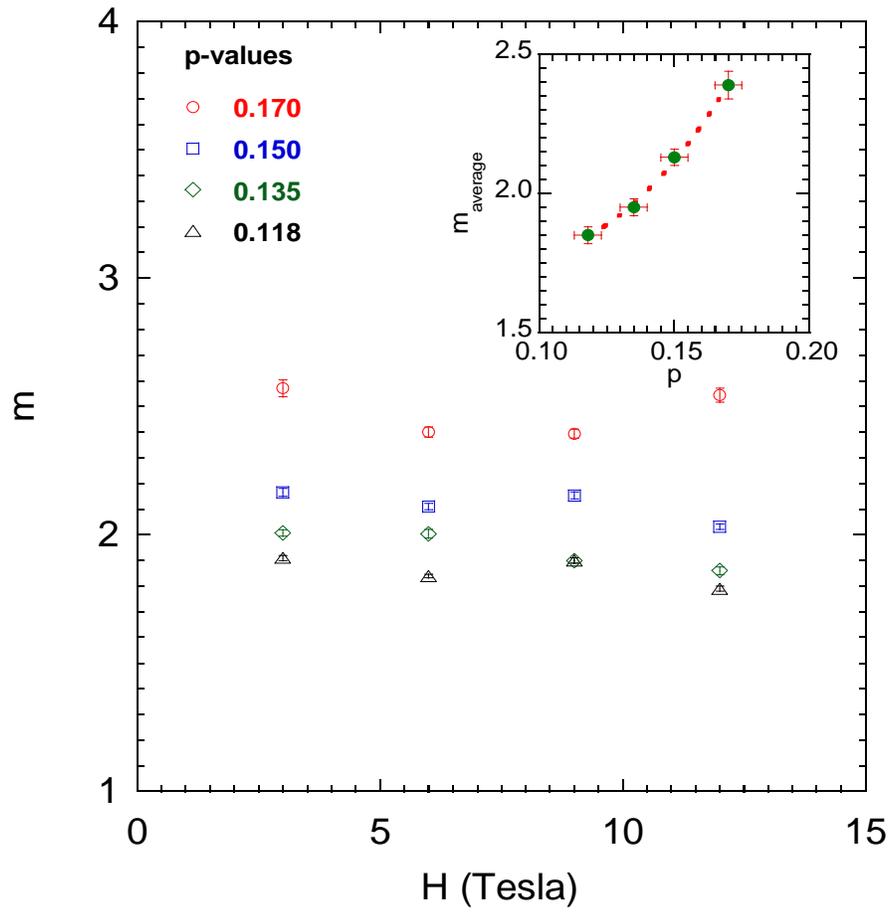

Figure 6

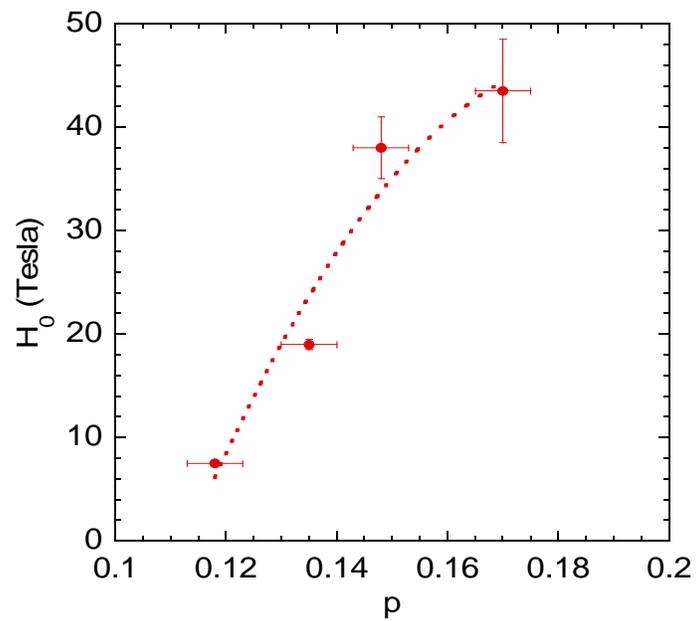

Figure 7

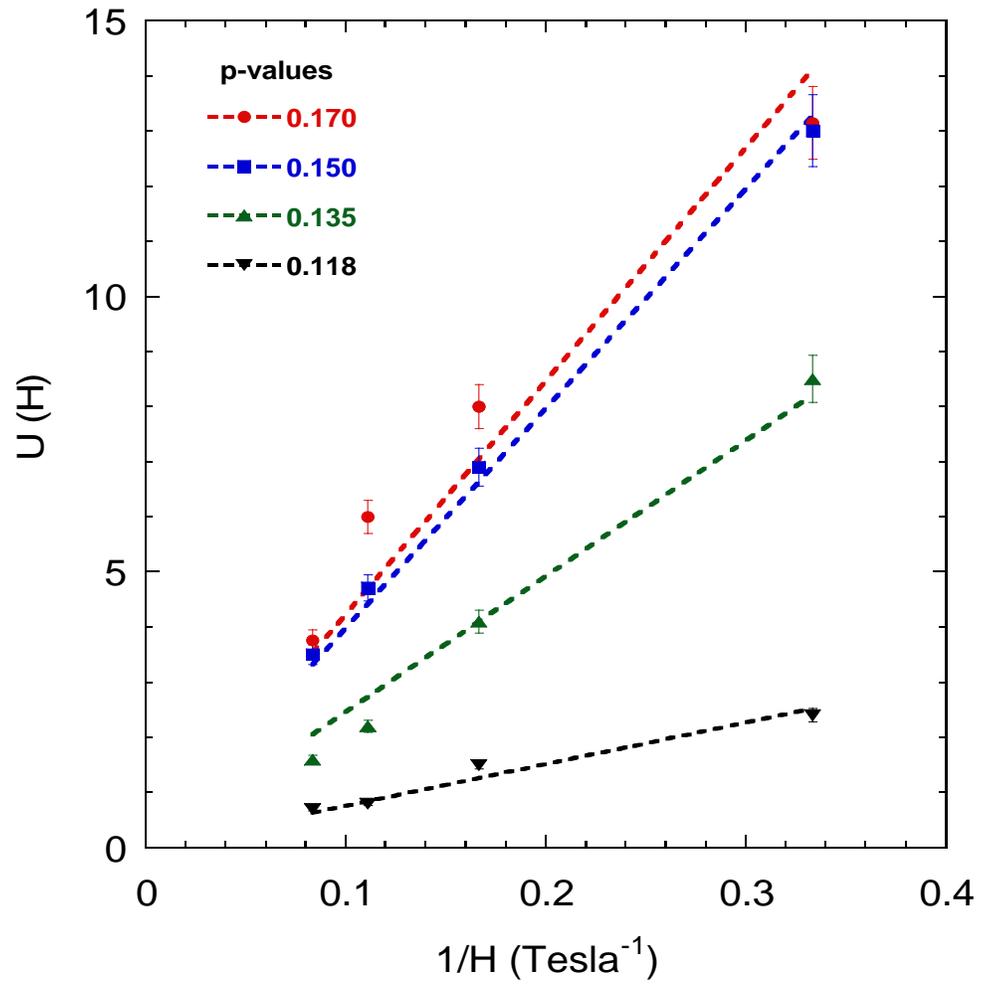